\documentclass[prb,twocolumn,footinbib, showpacs,showkeys]{revtex4}
\usepackage{graphicx}
\usepackage{epsfig}
\usepackage{dcolumn}
\usepackage{bm}
\usepackage{amsmath}
\usepackage{amsbsy}
\usepackage{amssymb}
\usepackage{color}
\usepackage{verbatim}

\newcommand{\bra}[1]{\ensuremath{\langle{#1}|\,}}
\newcommand{\ket}[1]{\ensuremath{\,|{#1}\rangle}}

\begin{document}

%{\huge{\color{red}RED} replaces {\color[rgb]{0.7,0.7,0.7} GREY.}
% comments in {\color{blue}BLUE}} \vspace{10mm}

%\large

\title{Twisted-light-induced optical transitions in semiconductors:
Free-carrier quantum kinetics}

\author{G.\ F.\ Quinteiro and P.\ I.\ Tamborenea}

\affiliation{Departamento de F\'{\i}sica and IFIBA,
Universidad de Buenos Aires, Ciudad Universitaria,
Pabell\'on I, 1428 Ciudad de Buenos Aires, Argentina}

\date{\today}

\begin{abstract}
We theoretically investigate the interband transitions and quantum
kinetics induced by
light carrying orbital angular momentum, or twisted light, in bulk
semiconductors.
We pose the problem in terms of the Heisenberg equations of motion
of the electron populations, and inter- and intra-band coherences.
Our theory extends the free-carrier Semiconductor Bloch Equations
to the case of photo-excitation by twisted light.
The theory is formulated using cylindrical coordinates, which are better
suited to describe the interaction with twisted light than the usual
cartesian coordinates used to study regular optical excitation.
We solve the equations of motion in the low excitation regime, and
obtain analytical expressions for the coherences and populations;
with these, we calculate the orbital angular momentum transferred
from the light to the electrons and the paramagnetic and diamagnetic
electric current densities.
\end{abstract}

\maketitle

% -----------------------------------------------------------------------------
% -----------------------------------------------------------------------------
% -----------------------------------------------------------------------------

\section{Introduction}

Since the seminal work by Allen {\it et al.} in 1992,\cite{all-bei-spr}
there has been a steady increase of interest in the theory, experiments,
and applications of light carrying orbital angular momentum (OAM),
or twisted light (TL).\cite{Andrews-01}
Studies in this area span several subfields of physics, such as research
on the classical/quantum properties
of TL,\cite{van-nie,all-pad,mol-tor-tor,dav-and-bab}
its generation,\cite{bar-tab}
and the interaction of TL with
atoms/molecules\cite{fri-etal,bab-etal,gar-etal,ara-etal}
and Bose-Einstein condensates.\cite{Sim_07}
At the same time, the interaction of general inhomogeneous light beams
with solids is becoming an active field of research
too.\cite{hes-kuh,ros-kuh,her-etal}

\begin{comment}
Light carrying orbital angular momentum (OAM), or twisted light (TL), is
attracting increasing attention.\cite{Andrews-01}
%
Also, the interaction of inhomogeneous light beams with solids is an
active field of research.\cite{Kuhn-01}
\end{comment}

Recently, we laid down the basic theoretical elements to study the
interaction of semiconductors and insulators with confined beams of
twisted light.\cite{qui-tam-09a}
We obtained the optical transition matrix elements of the TL-electron
interaction and studied the transfer of orbital angular momentum using
a simple, perturbative wave-function approach.
That approach was adequate as a first theoretical step, but it has a number
of limitations.
Being a single-particle theory, is has the drawback of not taking into
account the Pauli exclusion in the photo-excitation of multiple electrons,
and furthermore, it leaves out the electron-electron interaction effects.
Thus, a more complete theoretical treatment of the interband excitation
by twisted light of solids is called for.

In this paper we develop a set of ``twisted-light-generalized
semiconductor Bloch equations" (TL-SBE) from the Heisenberg equations of motion
of the populations and coherences of the photo-excited electrons.
This theory is valid for pulsed or CW twisted-light beams, and
takes fully into account the inhomogeneous profile of the beam, as well
as the transfer of momentum from the light to the electrons in the plane
perpendicular to the beam's propagation direction.
As long as excitonic phenomena are not targeted, the Coulomb interaction
does not play an essential role in the basic physics of band-to-band optical
transitions, and for that reason we will limit ourselves, for the time being,
to a free-carrier formulation of the theory.
From a practical point of view, we mention that the free-carrier theory
is already involved enough to merit a separate presentation, obviously as a
first step in a program that aims at obtaining and solving, first the
mean-field TL-SBE, and later the same equations with collision
terms.\cite{hau-jau}
Collision terms describe the scattering processes undergone by the
photoexcited electrons, namely, electron-electron, electron-phonon,
and electron-impurity scattering.
Collision terms in the relaxation-time-approximation can be added
straightforwardly to our theory in order to describe qualitatively those
scattering processes, and a numerical solution of the resulting
equations of motion would allow us to explore the influence of collisions
on the effects described here.
We leave this numerical treatment for future work, which will include,
besides collisions, the study of strong and pulsed TL excitation.
Finally, notice that while we concentrate here on bulk systems, our theory
can easily be applied to two-dimensional systems excited at normal incidence.

Usually, the optical excitation in bulk systems is theoretically
dealt with by assuming that the system is a cube, quantizing the
electrons using cartesian coordinates and taking, at the right
moment in the derivation, the limit of large system size.
For symmetry reasons, this method allows straightforward calculations in
the case of excitation with plane-wave light.
However, it leads to a
cumbersome formulation in the case of excitation by twisted light.
This is clearly so because the twisted light beam has an inherently
cylindrical nature.
In two previous works on the interacion of TL with quantum dots
\cite{qui-tam-09b} and quantum rings,\cite{qui-ber} the cylindrical
nature of the TL beams was handled conveniently by also using
cylindrical coordinates in the description of the electronic states.
In the theory presented here for bulk systems, we take advantage of this
simple but key idea.
We imagine the solid as a cylinder, quantize the electron states in
cylindrical coordinates, and finally take the limit of large system;
we rely on the fact that bulk properties are then independent of the
geometry of the solid.
We keep, naturally, the required microscopic
structure of the Bloch wave functions in order to characterize the
valence and conduction band states: the periodic parts of the Bloch
states are approximated by their values at zero crystal momentum, a
common practice known as effective-mass approximation.
The use of cylindrical rather than cartesian coordinates allowed us
to decouple the Heisenberg equations of motion according to values
of the electron angular momentum, which greatly reduces the
complexity of the problem.
Using these generalized TL-SBE we predict the kinetics of
electrons, show the occurrence of electric currents with complex
profiles, and demonstrate the transfer of OAM from the light beam to
the electrons.

The paper is organized as follows.
The TL vector potential and the system Hamiltonian are given in Sec.\
\ref{sec:system}.
Section \ref{sec:SBE} presents the generalized TL-SBE in terms of
cylindrical electron states, the partial decoupling of the equations of
motion, and the perturbative solution.
The electron quantum kinetics is analyzed with the help of the electric
current and transferred angular momentum in Sec.\ \ref{sec:kinetics}.
Conclusions are given in Sec.\ \ref{sec:conc}.

% -----------------------------------------------------------------------------
% -----------------------------------------------------------------------------
% -----------------------------------------------------------------------------

\section{system and twisted light}
\label{sec:system}

In this work we consider a direct-gap semiconductor or insulator and study
interband transitions caused by illumination with a beam of twisted light.
We assume that the light's frequency is such that mainly band-to-band
transitions occur so that exciton creation is unimportant.
Under this regime it is satisfactory to formulate our theory not
including the Coulomb interaction between carriers.
Thus, our theory can describe the electrons' kinetics, from
irradiation to a fraction of picoseconds---to avoid strong deviation
due to decoherence---in a large number of physical systems, from
semiconductors having band-gap of a fraction of an eV
(e.g.\ InSb with $0.23\, \mbox{eV}$) through several eV
(e.g.\ GaN  with $3.5 \, \mbox{eV}$), up to insulators
having larger gaps, provided that the frequency of the twisted field
is tuned above the energy bandgap; this requires the use of twisted
fields in the near-infrared to UV spectrum, which does not
constitute an experimental difficulty.
We are particularly interested in characterizing the transfer of
angular momentum between the TL and the electrons, and describing
what the electron distribution looks like as a result of the
photo-excitation.
Although several valence bands may be involved in interband optical
transitions, here we consider for simplicity a two-band model.
The generalization of the theory to the case with more than one valence
band involved is straightforward.

The vector potential of the TL beam in the Coulomb gauge is given by~\cite{jau}
\begin{eqnarray}
    \textbf{A}(\textbf{r},t)
&=& A_0 e^{i(q_z z- \omega t)}
    \,\left[ \boldsymbol{\epsilon}_{\pm} J_{l}(q_r r)
             e^{i l \phi} \mp \right. \nonumber \\
&&  \left.
    i \,\boldsymbol{\epsilon}_z
    \frac{q_r}{q_z} J_{l\pm 1}(q_r r)
    e^{i (l \pm 1) \phi} \right] + c.c. \,,
\label{eq:A}
\end{eqnarray}
with the polarization vectors given by
$\boldsymbol{\epsilon}_{\pm}
= \hat{x}\pm i\hat{y}
= e^{\pm i\phi}(\hat{r}\pm i\hat{\phi})$
and $c.c.$ denoting the complex conjugate.
In Eq.\ (\ref{eq:A}), the radial profile of the beam is given, for
concreteness, by Bessel functions, $J_{l}(q_r r)$ and $J_{l\pm
1}(q_r r)$ ---an alternative formulation would use Laguerre-Gaussian
modes instead~\cite{dav-and-bab}.

The light-matter interaction is described using the minimal-coupling
Hamiltonian, whose dominant contribution (for moderate field intensities)
comes from the linear term $\textbf{p} \cdot \textbf{A}$.
If the TL beam is such that $q_r < q_z$, which is verified for usual
sizes of the beam's waist, the largest coupling term comes from the
transverse component of $\textbf{A}(\textbf{r},t)$, and we only need
to consider
\begin{eqnarray}
    \textbf{A}(\textbf{r},t) &\simeq&
    \boldsymbol{\epsilon}_\sigma \, A_0 \,e^{i(q_z z-\omega t)}\,
    J_l(q_r r)\,  e^{il\phi} + c.c. \nonumber\\
    &=& \textbf{A}^{(+)}(\textbf{r},t) + \textbf{A}^{(-)}(\textbf{r},t)
    \,,
\label{eq:A_transverse}
\end{eqnarray}
with $\sigma = \pm$.

The lowest-order light-matter interaction Hamiltonian is
\begin{eqnarray}
h_{I} &=& - \frac{q}{m_e} \textbf{p} \cdot \textbf{A}(\textbf{r},t)\nonumber\\
      &=& - \frac{q}{m_e} \textbf{p} \cdot
            \left[ \textbf{A}^{(+)}(\textbf{r},t) +
                    \textbf{A}^{(-)}(\textbf{r},t)
            \right] \nonumber \\
      &=&  h_{I}^{(+)} + h_{I}^{(-)} \,,
\label{Eq_HI}
\end{eqnarray}
with $\mathbf p$ the momentum operator, and $m_e$ and $q$ the mass and
charge of the electron.
(Note that the $m_e$ appearing in this equation is the bare electron
mass and not the effective mass.) We emphasize that, unlike most
work in light-matter interaction, we must keep the spatial variation
of the field, in order to capture the relevant physics. Thus, Eq.\
(\ref{Eq_HI}) encodes all multipoles, as is clearly seen from the
Power-Zienau-Woolley transformation.\cite{Cohen}

The complete electronic Hamiltonian in second quantization for the
general multi-band case is
\begin{eqnarray*}
    {\cal H}
&=& \sum_{b \alpha} \varepsilon_{b \alpha }
    \, a^\dag_{b \alpha }  a_{b \alpha } +
    \sum_{b \alpha  , b' \alpha'}
    \bra{b' \alpha'} h_I \ket{b \alpha}
    \, a^\dag_{b' \alpha' } a_{b \alpha }
\end{eqnarray*}
where $b, b'$ denote energy bands, like heavy hole, light hole,
conduction, etc., $\alpha$ is a collective index of quantum
numbers appropriate for the problem at hand, and $a^\dag/a$ are
creation/annihilation operators.

%------------------------------------------------------------------------------
%------------------------------------------------------------------------------
%------------------------------------------------------------------------------

\section{Free-carrier Semiconductor Bloch equations}
\label{sec:SBE}

%------------------------------------------------------------------------------
%------------------------------------------------------------------------------
\subsection{General equations of motion}

Let us consider the operator $\hat{\rho}_{b' \alpha' ,b \alpha } =
a^\dag_{b' \alpha' } a_{b \alpha}$.
The equation of motion for this operator in the Heisenberg picture is
\begin{eqnarray}
    i\hbar \frac{d}{dt}\hat{\rho}_{b' \alpha',b \alpha}
&=& [\hat{\rho}_{b' \alpha',b \alpha}, {\cal H}] \,.
\label{eq:em_general}
\end{eqnarray}
For concreteness, we take one type of circularly polarized light,
either $\sigma_+$ or $\sigma_-$.
While, say, $\sigma_+$ light connects both light holes and heavy
holes to conduction band states, these two processes remain
independent of each other during the evolution under the Hamiltonian
that we are considering.
Then, for circularly polarized light, we can accurately describe the
electron kinetics under the optical excitation within a two band
model.
We now specialize Eq.\ (\ref{eq:em_general}) to a two-band case by
considering the evolution of the three types of operators
$\hat{\rho}_{c \alpha', c \alpha}$, $\hat{\rho}_{v \alpha', v
\alpha}$, and $\hat{\rho}_{v \alpha', c \alpha}$,
where $v$ ($c$) stands for the chosen valence (conduction) band.
After expanding the commutators and assuming that the interaction $h_I$
connects only valence- to conduction-band states, the equations of motion
become
%
%\begin{widetext}
\begin{eqnarray}
    i\hbar \frac{d}{dt}\hat{\rho}_{c \alpha', c \alpha}
&=& \Delta_{c \alpha, c \alpha' }
    \, \hat{\rho}_{c \alpha', c \alpha} +\nonumber \\
&&  \hspace{-20mm}
    \sum_{\alpha_1}
    \bra{c \alpha} h_I \ket{v \alpha_1}
    \hat{\rho}_{c \alpha', v \alpha_1} -
    \bra{v \alpha_1} h_I \ket{c \alpha'}
    \hat{\rho}_{v \alpha_1, c \alpha},
\label{eq:rho_c} \\
    i\hbar \frac{d}{dt}\hat{\rho}_{v \alpha', v \alpha}
&=& \Delta_{v \alpha, v \alpha' }
    \, \hat{\rho}_{v \alpha', v \alpha} +\nonumber \\
&&  \hspace{-20mm}
    \sum_{\alpha_1}
    \bra{v \alpha} h_I \ket{c \alpha_1}
    \hat{\rho}_{v \alpha', c \alpha_1} -
    \bra{c \alpha_1} h_I \ket{v \alpha'}
    \hat{\rho}_{c \alpha_1, v \alpha},
\label{eq:rho_v} \\
    i\hbar \frac{d}{dt}\hat{\rho}_{v \alpha', c \alpha}
&=& \Delta_{c \alpha, v \alpha' }
    \, \hat{\rho}_{v \alpha', c \alpha} +\nonumber \\
&&  \hspace{-20mm}
    \sum_{\alpha_1}
    \bra{c \alpha} h_I \ket{v \alpha_1}
    \hat{\rho}_{v \alpha', v \alpha_1} -
    \bra{c \alpha_1} h_I \ket{v \alpha'}
    \hat{\rho}_{c \alpha_1, c \alpha} \,,
\label{eq:rho_vc}
\end{eqnarray}
%\end{widetext}
%
where $\Delta_{b \alpha, b' \alpha' } = \varepsilon_{b \alpha } -
\varepsilon_{b' \alpha' }$.

In what follows we work with the equations of motion of the
expectation values of the operators $\hat{\rho}$:
\begin{eqnarray}
\rho_{c, \alpha' \alpha} =
   \langle \hat{\rho}_{c \alpha', c \alpha} \rangle, \nonumber \\
\rho_{v, \alpha' \alpha}   =
   \langle \hat{\rho}_{v \alpha', v \alpha} \rangle, \nonumber \\
\rho_{v \alpha', c \alpha} =
   \langle \hat{\rho}_{v \alpha', c \alpha} \rangle \,,
\end{eqnarray}
where the average $\langle \ldots \rangle$ is taken over the initial
state of the material.
These expectation values represent populations when they have
repeated indices and quantum coherences when they are off-diagonal
matrix elements.
Notice that in Eqs.\ (\ref{eq:rho_c}-\ref{eq:rho_vc}) we keep the
{\em intraband} coherences, which are essential in the TL excitation
process.
These coherences are usually left out of the theory when the {\it
vertical transition} approximation is made.

%------------------------------------------------------------------------------
%------------------------------------------------------------------------------

\subsection{Electronic states in cylindrical coordinates}

The semiconductor Bloch equations for bulk systems are usually
formulated in the basis of electronic Bloch states given by
\begin{equation}\label{Eq_funct}
    \psi_{b \textbf{k}}(\textbf{r}) =
    \langle \textbf{r} \ket{b \textbf{k}} = \frac{1}{L^{3/2}}
    e^{i \textbf{k} \cdot \textbf{r}} u_{b \mathbf k}(\textbf{r})\,,
\end{equation}
where $b$ is the band index, $\mathbf k$ is the crystal momentum,
and $L$ is the linear size of the system.
In principle, this basis set could be used as well in our treatment of
TL-excited systems, but one finds that it is not a convenient choice.
To see this, let us recall the interband matrix element of the
TL-matter interaction Hamiltonian given in Ref.\ [\onlinecite{qui-tam-09a}]:
\begin{eqnarray*}
    \bra{c {\bf{k}}^\prime} h_I^{(+)} \ket{v {\bf{k}}}
&=& -(-i)\,^l\,\frac{q}{m_e} \,\frac{A_0(t)}{L}
    \,\frac{1}{q_r} \,\delta_{\kappa_r, \, q_r}
    \,\delta_{\kappa_z,\,q_z} \\
&&  e^{i\,\theta \,l}
    \,\left(\boldsymbol{\epsilon}_\sigma \cdot {\bf p}_
    {c v}\right) \, e^{-i\,\omega\, t}
\end{eqnarray*}
where $\boldsymbol{\kappa} = \mathbf{k}' -\mathbf{k}$ has an azimuthal
angle $\theta$ and its projection in the x-y plane has
length $\kappa_r$, and $\mathbf p_{b' b} = (1/a^3) \int_{c} d^3x \,
u_{b'}^*(\mathbf r) \, \hat{ \mathbf p} \, u_{b }(\mathbf r) $.
After inserting this matrix element (and its complex conjugate) into
Eqs.\ (\ref{eq:rho_c})-(\ref{eq:rho_vc}), one immediately realizes
that a single state in the valence (conduction) band is connected
to a multitude of states in the conduction (valence) band having any
angle $\theta$; pictorially, this has been represented by us in
Ref.\ [\onlinecite{qui-tam-09a}] by a cone-like excitation in
momentum representation.
Thus, the equations are almost completely coupled---with the
exception of the $z$ component---and even the perturbation-theory
solution looks complicated and hard to interpret.

As anticipated in the introduction, there are compelling reasons to
adopt a different basis set for the electrons in the solid:
$i)$ the symmetry exhibited by the vector potential
Eq.\ (\ref{eq:A_transverse});
$ii)$ (bulk) properties are the same for a box- or cylinder-shaped solid,
in the large-system limit;
and $iii)$ optical excitations are well described in a band-edge or
effective mass approximation.
Therefore, we adopt cylindrical states to treat the electrons;
their wave functions and energies are (see Appendix \ref{sec:cylinder_states})
\begin{eqnarray}
    \psi_{b \mathbf k m}(\mathbf r)
&=& {\cal N} J_m(k_r \,r)
    e^{i m \phi} e^{i k_z z} u_{b}(\mathbf r) \,,
    \nonumber \\
    \varepsilon_{b \mathbf k m}
&=& \frac{\hbar^2}{2 m_b^*}
    \left( k_r^2 + k_z^2 \right) + \delta_{b c} \, E_g \,,
\label{eq:electron_basis}
\end{eqnarray}
where $\mathbf k$ represents the quantum numbers $\{k_z,k_r\}$, with
$k_z=2 \pi n/H$ and $k_r = r_{m s}/L$.
\footnote{We formerly used the symbol $\parallel$ to represent the
in-plane component.}
$H$ and $L$ are the height and radius of the cylinder, respectively,
$ r_{m s}$ is the $s$th root of the Bessel function of order $m$,
the normalization constant ${\cal N}$ depends on $\{m,s\}$,
and $n$ is an integer.
In this basis set, the light-matter interaction matrix elements read
\begin{eqnarray} \label{eq:Matrix_Element}
    \bra{c \mathbf k' m'} h_I^{(+)} \ket{v \mathbf k m}
&=& \xi_{c k_r' m', v k_r m} \, e^{-i\omega t}\, \times \nonumber \\
&&  \delta_{l, m' - m } \delta_{q_z, k_z' - k_z } \nonumber \\
    \bra{v \mathbf k' m'} h_I^{(-)} \ket{c \mathbf k m}
&=& \xi_{c k_r m, v k_r' m'}^* \, e^{i\omega t}\, \times \nonumber \\
&&  \delta_{-l, m' - m } \delta_{-q_z, k_z' - k_z }
\,,
\end{eqnarray}
where $\xi_{b' k_r' m, b k_r m} = - \frac{q}{m_e} A_0 (\mathbf
p_{b'b} \cdot \boldsymbol{\epsilon}_\sigma) \, {\cal N}' {\cal N}
\int_0^{L} d r \, r \,
\times \\
J_l(q_r\,r)  J_{l+m}(k_r'\,r) \,J_m(k_r\,r)$.

In order to derive the TL-generalized SBE, we specialize Eqs.\
(\ref{eq:rho_c}-\ref{eq:rho_vc}) to cylindrical states, for which
the composite quantum index is $\alpha=\{k_z, k_r, m \}$.
The plan is to write the equations of motion only for those
components of $\rho$ which are effectively coupled among themselves.
In this way, we separate the evolution on the whole Hilbert space into
that of dynamically uncoupled subspaces.
Thus, we proceed by first writing down the equation of motion of a
population or an intraband coherence, say, of the valence band.
It can be seen that if the semiconductor is initially in its
non-interacting ground state
(it is universally accepted in the SBE literature to assume the absence
of Coulomb correlations in the unexcited material) we only need to consider
cases with $\{ k_z' = k_z, m'=m \}$, since all other coherences remain
zero at all times.\cite{qui-ber}
Thus we write
\begin{widetext}
\begin{eqnarray}
i\hbar \frac{d}{dt} \rho_{v k_z k_r' m, \, k_z k_r m}
&=&
    \Delta_{v k_z k_r m, \, k_z k_r' m}
    \, {\rho}_{v k_z k_r' m, \, k_z k_r m} + \nonumber \\
&&  \hspace{-25mm}
    e^{i\omega t} \sum_{k_r''} \,
    \xi_{c k_r'' m+l, \, v k_r m}^* \,
    {\rho}_{v k_z k_r' m, \, c k_z+q_z k_r'' m+l} -
    e^{-i\omega t} \sum_{k_r''}
    \xi_{c k_r'' m+l, \, v k_r m}\,
    {\rho}_{c k_z+q_z k_r'' m+l, \, v k_z k_r m} \, .
\label{eq:TL-SBE_v}
\end{eqnarray}
We see that $\rho_{v k_z k_r' m, \, k_z k_r m}$ gets coupled to an
interband coherence ${\rho}_{v k_z k_r' m, \, c k_z+q_z k_r'' m+l}$
and ${\rho}_{c k_z + q_z k_r'' m+l, \, v k_z k_r m}$, with all
values $k_r''$ of the radial quantum number, but with just $\{k_z, m
\}$ and $\{k_z+q_z, m+l \}$ for the other two quantum numbers.
The Heisenberg equation for these interband coherences is
\begin{eqnarray}
i\hbar \frac{d}{dt} {\rho}_{v k_z k_r' m, \, c k_z+q_z k_r m+l} &=&
    \Delta_{c k_z+q_z k_r m+l, \, v k_z k_r' m }
    \, {\rho}_{v k_z k_r' m, \, c k_z+q_z k_r m+l} +\nonumber \\
&&  \hspace{-25mm}
    e^{-i\omega t}
    \sum_{k_r''}
    \xi_{c k_r m+l, \, v k_r'' m} \,
    {\rho}_{v k_z k_r' m, \, k_z k_r'' m} -
    e^{-i\omega t}  \sum_{k_r''}
    \xi_{c k_r'' m+l, \, v k_r' m} \,
    {\rho}_{c k_z+q_z k_r'' m+l, \, k_z+q_z k_r m+l} \,,
\label{eq:TL-SBE_vc}
\end{eqnarray}
with $\rho_{v \alpha', c \alpha} = \rho_{c \alpha, v \alpha'}^*$.
Inspection of this equation reveals that the interband coherence is
coupled back to the initial valence-band coherence
[Eq.\ (\ref{eq:TL-SBE_v})] and additionally to a conduction-band coherence
or population, whose equation of motion is
\begin{eqnarray}
i\hbar \frac{d}{dt} {\rho}_{c k_z+q_z k_r'
m+l, k_z+q_z k_r m+l} &=&
    \Delta_{c k_z+q_z k_r m+l, k_z+q_z k_r' m+l}
    \,{\rho}_{c k_z+q_z k_r' m+l, k_z+q_z k_r m+l} +
    \nonumber \\
&&  \hspace{-50mm}
    e^{-i\omega t}
    \sum_{k_r''}
    \xi_{c k_z+q_z k_r' m+l, v k_z k_r'' m}
    \,{\rho}_{c k_z+q_z k_r' m+l, v k_z k_r'' m} -
    e^{i\omega t}
    \sum_{k_r''}
    \xi_{c k_z+q_z k_r m+l, v k_z k_r'' m}^*
    \,{\rho}_{v k_z k_r'' m, c k_z+q_z k_r m+l} \,.
\label{eq:TL-SBE_c}
\end{eqnarray}
\end{widetext}
This equation couples the population or intraband coherence
${\rho}_{c k_z+q_z k_r' m+l, k_z+q_z k_r m+l}$ to interband coherences
which evolve according to Eq.\ (\ref{eq:TL-SBE_vc}).
It is clear that the system of equations is closed in the subspaces
of fixed $\{ k_z, m \}$ and $\{ k_z + q_z, m+l\}$, and the complexity of the
problem has been drastically reduced, compared to the system of
equations found in the case of Bloch states for a cubic bulk material.
We may say that our procedure is equivalent to a block diagonalization.
At this stage, the problem is highly tractable by computational techniques,
since the only unconstrained variable is $k_r$.

%------------------------------------------------------------------------------
\subsection{Low-excitation regime}

Equations (\ref{eq:TL-SBE_v})-(\ref{eq:TL-SBE_c}) in all their generality
are not amenable to analytical treatment.
However, in the case of low photo-excitation, an analytical perturbative
approach is possible and gives us the basic physical insight that we
are looking for.
We now pursue this approach, but work initially with the system of equations
(\ref{eq:rho_c})-(\ref{eq:rho_vc}) instead of
(\ref{eq:TL-SBE_v})-(\ref{eq:TL-SBE_c}), since the former are more general
and also notationally simpler than the latter.
We solve the system to lowest order in the vector potential, that is, we
first solve Eq.\ (\ref{eq:rho_vc}) assuming that the zeroth-order intraband
elements are
$\rho_{v, \alpha' \alpha} = \delta_{\alpha',\alpha}$ and
$\rho_{c, \alpha' \alpha} = 0$,
and then solve Eqs.\ (\ref{eq:rho_c}) and (\ref{eq:rho_v}) using the
first-order solution of (\ref{eq:rho_vc}).
The equation of motion for the interband polarization
$\rho_{v \alpha', c \alpha}$ becomes
\begin{eqnarray} \label{Eq:Interband}
    \left[ i\hbar \frac{d}{dt}
          - (\varepsilon_{c \alpha } - \varepsilon_{v \alpha' })
    \right] \rho_{v \alpha', c \alpha}^{(1)}
&=& \bra{c \alpha} h_I \ket{v \alpha'}\,.
\end{eqnarray}
For a monochromatic electromagnetic field turned on at $t=0$
the solution reads
\begin{eqnarray}
\label{Eq_lowestO_pol}
    \rho_{v \alpha', c \alpha}^{(1)}(t)
&=& Y_{c \alpha, v \alpha'}(t) \,
    \bra{c \alpha} h_{I}^{(+)} \ket{v \alpha'} \,,
\end{eqnarray}
with
\begin{eqnarray*}
    Y_{c \alpha, v \alpha'}(t)
&=& - \frac{1-e^{-i [(\varepsilon_{c \alpha }-
    \varepsilon_{v \alpha'})-\hbar\omega]\, t / \hbar}}
    {(\varepsilon_{c \alpha }-
    \varepsilon_{v \alpha'})-\hbar\omega} \,.
\end{eqnarray*}
\begin{widetext}
\noindent Inserting $\rho_{v \alpha', c \alpha}^{(1)}$ in the
equations for the intra-band coherence, Eqs.\ (\ref{eq:rho_c}) and
(\ref{eq:rho_v}), we obtain
\begin{eqnarray}
\rho_{c, \alpha', \alpha}^{(2)}(t)
&=& -
    \frac{i}{\hbar}
    e^{-i(\varepsilon_{c \alpha}- \varepsilon_{c \alpha'})t/\hbar}
    \sum_{\alpha_1}
    \bra{c \alpha} h_{I}^{(+)} \ket{v \alpha_1}
    \bra{v \alpha_1} h_{I}^{(-)} \ket{c \alpha'}
    \times \\
&&  \int_{0}^{t} dt'\,
    e^{i(\varepsilon_{c \alpha}-
    \varepsilon_{c \alpha'}) t'/ \hbar} \,
    \left[ Y_{c \alpha', v \alpha_1}^*(t') -
           Y_{c \alpha, v \alpha_1}(t')
    \right], \nonumber \\
\rho_{v, \alpha', \alpha}^{(2)}(t)
&=& \delta_{\alpha', \alpha} -
    \frac{i}{\hbar}
    e^{-i(\varepsilon_{v \alpha}- \varepsilon_{v \alpha'})t/\hbar}
    \sum_{\alpha_1}
    \bra{v \alpha} h_{I}^{(-)} \ket{c \alpha_1}
    \bra{c \alpha_1} h_{I}^{(+)} \ket{v\alpha'}
    \times \\
&&  \int_{0}^{t} dt'\,
    e^{i(\varepsilon_{v \alpha}-
    \varepsilon_{v \alpha'}) t'/ \hbar} \,
    \left[ Y_{c \alpha_1, v \alpha'}(t') -
           Y_{c \alpha_1, v \alpha}^*(t')
    \right].
\label{Eq_lowestO_pop}
\end{eqnarray}
For example, the conduction-band populations reduce to
\begin{equation}
\rho_{c, \alpha, \alpha}^{(2)}(t) = \frac{2}{\hbar}
    \sum_{\alpha_1}
    \frac{|\bra{c \alpha} h_{I}^{(+)} \ket{v \alpha_1}|^2}
         {(\varepsilon_{c \alpha}
         - \varepsilon_{v \alpha_1}
         - \hbar \omega)^2}
    \left\{
        1 -
        \cos
        \left[
            (\varepsilon_{c \alpha}
            - \varepsilon_{v \alpha_1}
            - \hbar \omega) t / \hbar
        \right]
    \right\}
    \,,
\end{equation}
where one sees that their time evolution is slow, with frequencies
close to the detuning.
%
%Also, they are real as expected, and energy denominators appear as usual in
%perturbation theory.
%
As for the products of matrix elements, in our problem,
with $\alpha=\{k_z,k_r,m\}$, a simple calculation using
Eqs.\ (\ref{eq:Matrix_Element}) yields
\begin{equation}
    \bra{c \alpha} h_{I}^{(+)} \ket{v \alpha_1}
    \bra{v \alpha_1} h_{I}^{(-)} \ket{c \alpha'} =
    \xi_{c k_r m, v k_{1r} m_1}
    \xi_{c k_r' m, v k_{1r} m_1}^*
    \delta_{m',m} \delta_{k_z,k_z'}
    \delta_{m_1,m-l} \delta_{k_{1z}, k_z - q_z} \nonumber
\end{equation}
\begin{equation}
    \bra{v \alpha} h_{I}^{(-)} \ket{c \alpha_1}
    \bra{c \alpha_1} h_{I}^{(+)} \ket{v\alpha'} =
    \xi_{c k_{1r} m_1, v k_r m}^*
    \xi_{c k_{1r} m_1, v k_r' m}
    \delta_{m',m} \delta_{k_z,k_z'}
    \delta_{m_1,m+l} \delta_{k_{1z}, k_z + q_z}
\, ,
\label{eq:prod_mat_elem}
\end{equation}
\end{widetext}
%
%
%\begin{eqnarray*}
%    \bra{c \alpha} h_{I}^{(+)} \ket{v \alpha_1}
%    \bra{v \alpha_1} h_{I}^{(-)} \ket{c \alpha'}
%&=& \xi_{c k_r m, v k_{1r} m_1} \\
%&&  \hspace{-30mm}
%    \xi_{c k_r' m, v k_{1r} m_1}^*
%    \delta_{m',m} \delta_{k_z,k_z'}
%    \delta_{m_1,m-l} \delta_{k_{1z}, k_z - q_z} \\
%----
%    \bra{v \alpha} h_{I}^{(-)} \ket{c \alpha_1}
%    \bra{c \alpha_1} h_{I}^{(+)} \ket{v\alpha'}
%&=& \xi_{c k_{1r} m_1, v k_r m}^* \nonumber \\
%&&  \hspace{-30mm}
%    \xi_{c k_{1r} m_1, v k_r' m}
%    \delta_{m',m} \delta_{k_z,k_z'}
%    \delta_{m_1,m+l} \delta_{k_{1z}, k_z + q_z}
%\, ,
%\end{eqnarray*}
%
in agreement with the decoupling of
Eqs.\ (\ref{eq:TL-SBE_v})-(\ref{eq:TL-SBE_c}), showing that the
second-order process involves an intermediate state $\alpha_1$ which can
only differ from the initial state by $\pm l$ and $\pm q_z$ in the
quantum numbers $m$ and $k_z$, respectively.
Note that the interband coherence is of order
${\cal O}[\mathbf A(\mathbf r,t)^1]$,
while the populations and intraband coherences are of order
${\cal O}[\mathbf A(\mathbf r,t)^2]$,
as indicated with superscripts.
Finally, the time behavior of each component is clearly discernible:
while the interband coherence oscillates at the frequency of the TL field,
the populations and intraband coherences do it typically at terahertz
frequencies associated with interband Rabi flops and intraband energy
differences.

% -----------------------------------------------------------------------------
% -----------------------------------------------------------------------------

\section{Electron quantum kinetics}
\label{sec:kinetics}

The solutions to Eqs.\ (\ref{eq:TL-SBE_v})-(\ref{eq:TL-SBE_c}) are the
building blocks for constructing mean values of observables of interest.
In the standard theory of optical transitions, where
the light is assumed to be a plane wave and
the dipole approximation is made,
once the time- and momentum-dependent density matrix is obtained, one
calculates the macroscopic optical polarization and from it, for example,
the optical susceptibility.\cite{hau-koc}
Under those assumptions, the macroscopic polarization is just a spatially
uniform, time-dependent, function.

By contrast, if the inhomogeneities of the field are taken into account
(e.g.\ finite beam waist and oscillatory dependence in the propagation
direction), the electronic variables acquire an intricate space
dependence.
The excitation of solids by TL beams also produces a space-dependent
carrier kinetics which requires local variables for its description.
In order to visualize the pattern of motion of the photo-excited electrons,
we calculate in this Section the spatially inhomogeneous electric
current density.
Another useful variable, the transferred angular momentum, is instead
a global magnitude that characterizes the TL-material interaction.
Here we calculate their dynamics up to second order in the field amplitude.

In the calculations that follow, we will study separately the contributions
to the angular momentum and the electric current made by the interband
coherences, on the one hand, and by the populations and intraband coherences,
on the other.
This separation is conceptually useful because the interband coherence
contributions are fast (femtosecond) oscillations around a null value of
the current or angular momentum,
analogously to what happens with the interband polarization,
while the population or intraband-coherence contributions come from
slower (picosecond) processes in which a net transfer of momentum
from light to matter occurs.
The latter are related to the photon-drag effect,\cite{gib-wal,gib-mon}
which is now generalized to incorporate a {\it rotational} drag in the plane
perpendicular to the propagation direction, due to the ``slow" transfer of
angular momentum.
Furthermore, as we will see below, the lowest-order contributions for
interband and intraband processes are of first and second order in
the light field, respectively.

An investigation of the transfer of orbital angular momentum and the
generation of paramagnetic currents in semiconductors and insulators
was presented by us in a previous publication.\cite{qui-tam-09a}
In that study we employed a simple wave-function approach, which was
limited to describe the single-particle dynamics.
Our current theoretical analysis uses the formalism of the Heisenberg
equations of motion for populations and coherences,
which fully accounts for the Pauli exclusion in the multi-electron
excitation process, and has the advantage that it can be extended to
include electron-electron interaction.
In what follows, we analyze with this tool the transfer of orbital angular
mometum and the generation of electric currents, including the diamagnetic
term---missing in our previous study.

% -----------------------------------------------------------------------------
% -----------------------------------------------------------------------------
\subsection{Transfer of angular momentum}
\label{Sec:OAM}

Since the TL beam carries angular momentum around the $z$-axis, we focus on
the corresponding quantity for electrons,
\begin{eqnarray}
\label{eq:Lz}
  \hat{L}_z (t) &=& \sum_{b' b \alpha \alpha'}
    \bra{b' \alpha'} \hat{l}_z \ket{b \alpha}
    \, a^\dag_{b' \alpha'} (t) \, a_{b \alpha}(t) \,,
\end{eqnarray}
where $\hat{l}_z= - i \hbar \partial_\phi$.
We split the matrix-element integral
$\bra{b' \alpha'} \hat{l}_z\ket{b \alpha}$
into an integral on the unit cell and a sum over lattice sites.
Naturally, care must be taken when operating with $-i \hbar
\partial_\phi$ on the envelope, $\Phi_{\mathbf k m}(\mathbf r) = {\cal
N} J_m(k_r \,r) e^{i m \phi} e^{i k_z z}$, and on the periodic,
$u_{b}(\mathbf r)$, parts of the wave function $\psi_{b \mathbf k
m}(\mathbf r)$.
We obtain
\begin{eqnarray*}
    \bra{b' \mathbf k' m'} \hat{l}_z \ket{b \mathbf k m}
&=& \delta_{k_z',k_z} \delta_{mm'} \delta_{k_r',k_r}
    (\delta_{bb'} \hbar m  + \ell_{z,b' b}) + \\
&&  \hspace{-0mm}
    (1 - \delta_{bb'}) \, {\cal L}_{b'\mathbf k' m', b\mathbf k m}
    \,,
\end{eqnarray*}
where
\begin{equation}
\label{eq:lcal}
 {\cal L}_{b'\alpha', b\alpha} = \int d^3r \,
    \Phi_{\alpha'}^*(\mathbf r) \,
    \Phi_{\alpha}(\mathbf r) \,
    \left.\textbf{r} \times \mathbf p_{b' b} \right|_z,
\end{equation}
and this integral is over the whole crystal.
We note that the quantity $\ell_{z,b' b}$ does not depend on the OAM of
light, and so we disregard it from now on.
Next, we split the angular momentum into interband (coherence)
and intraband (population and coherence) contributions, and use
the perturbation-theory solutions of
Eqs.\ (\ref{Eq_lowestO_pol})-(\ref{Eq_lowestO_pop}).

% -----------------------------------------------------------------------------
{\it Interband-coherence contribution:}
After taking the mean value of the angular momentum operator [written
in second-quantization notation in Eq.\ (\ref{eq:Lz})] on the initial state,
we identify the component of the electronic angular momentum driven by
the interband coherence as
\begin{eqnarray} \label{eq:Lzcoh}
  L_z^{(coh)}(t)
  &=&
  \!\!\sum_{\mathbf k' m' \mathbf k m}
  \!\!\!2\Re\left[
    {\cal L}_{v \mathbf k' m', c \mathbf k m}
    \, {\rho}_{v \mathbf k' m', c \mathbf k m}^{(1)}(t)
  \right] .
\end{eqnarray}
We note that ${\cal L}_{v \mathbf k' m', c \mathbf k m}$
contains the factor
$\delta_{k_z',k_z}$, while
${\rho}_{v \mathbf k' m', c \mathbf k m}^{(1)}(t)$
contains the factor $\delta_{k_z,k_z' + q_z}$.
This mismatch, which renders a vanishing $L_z^{(coh)}$, comes from
dropping the dipole approximation in the calculation of the interband
coherence, and, at the same time, considering that the solid
is infinite in extent [in Eq.\ (\ref{eq:lcal})].
To be consistent, one needs to consider the system as a thin slice
of semiconductor perpendicular to the $z$-axis, having a width much
smaller than the wavelength of the light.
As a consequence, if the slice is located at $z_0$,
$\delta_{k_z',k_z}$ is replaced by $\exp[{i (k_z-k_z') z_0}]$
and we obtain
\begin{eqnarray*}
    {\cal L}_{b'\mathbf k' m', b\mathbf k m}
&=& i\,\pi\, f_{k_r' m', k_r m}\,e^{i (k_z-k_z') z_0} \times \\
&&  \left[
    \delta_{-1, m-m'} p_{-,b' b} -
    \delta_{1, m-m'} p_{+,b' b}
     \right]
    \,,
\end{eqnarray*}
with
$f_{k_r'm',k_rm} = {\cal N}' {\cal N} \int dr \,r^2\,
J_{m'}(k_r'\,r) \, J_m(k_r\,r)$,
and
$p_{\pm,b' b}= \hat{x}\cdot \mathbf p_{b' b} \pm
i \hat{y}\cdot \mathbf p_{b' b}$.
Then, we may succinctly state that
${\cal L}_{b'\mathbf k' m', b\mathbf k m} \propto \delta_{\pm 1, m-m'}$.
Inspection of the coherence, Eq.\ (\ref{Eq_lowestO_pol}),
and matrix elements, Eq.\ (\ref{eq:Matrix_Element}), shows that
${\rho}_{v \mathbf k' m', c \mathbf k m}^{(1)} \propto \delta_{l,m-m'}$.
Thus, we conclude that at this level there is a transfer of angular
momentum back and forth between the light beam and the electrons
if and only if $|l|= 1$.
We emphasize that, on time-average, there is no net transfer of angular
momentum to the material system, unless the temporal shape
of the electromagnetic pulse is asymmetric.\cite{Moska-01}

% -----------------------------------------------------------------------------
{\it Population and intraband-coherence contribution:}
The component of angular momentum driven by the populations and
intraband coherences reads
\begin{eqnarray*}
    L_z^{(pop)}(t)
&=& \sum_{\mathbf k m} \hbar m
    \, \left[{\rho}_{v \mathbf k m, \mathbf k m}^{(2)}(t) +
    {\rho}_{c \mathbf k m, \mathbf k m}^{(2)}(t)\right]\,.
\end{eqnarray*}
In order to correctly interpret this expression, we recall that the
TL photo-excitation process connects the valence-band subspace of
fixed $\{k_z, m\}$ with the conduction-band subspace of fixed
$\{k_z+q_z, m+l\}$;
thus, an imbalance population in the conduction band is produced.
This asymmetry between populations in both bands
with respect to the quantum number $m$ brings about a net angular momentum
acquired by the electrons, which we refer to as rotational photon-drag.
In contrast to $L_z^{(coh)}(t)$, $L_z^{(pop)}(t)$ has no restrictions
on the values of $l$ that cause a transfer of angular momentum, and its
time average yields a non-zero value.

% -----------------------------------------------------------------------------
% -----------------------------------------------------------------------------
\subsection{Induced currents}

Next we will obtain the photo-induced currents produced by the
irradiation with TL.
The general expressions of the electric current in second quantization notation
are as follows.
The more standard, paramagnetic, current density is given by $
\hat{\mathbf j}^{(p)}(x,t) = -{i q \hbar}/({2\,m_e}) \lim_{x^\prime
\rightarrow x} (\nabla-\nabla^\prime) \psi^\dag(x^\prime, t) \,
\psi(x, t)$ and the diamagnetic term is given by $\hat{\mathbf
j}^{(d)}(x,t) =-q/m_e \mathbf A(\mathbf r, t) \, \psi^\dag(x, t) \,
\psi(x, t)$ \cite{Ibach}.
We apply these expressions to our problem, and after some
algebraic manipulation, we obtain, for the paramagnetic term:
\begin{eqnarray}
     \hat{\mathbf j}^{(p)}(\mathbf r,t)
&=& -i\,\frac{q\,\hbar}{2\,m_e}
        \sum_{\scriptsize
                \begin{array}{c}
                b' \mathbf k' m'  \\
                b  \mathbf k  m
              \end{array}
             }
        \left[
            \psi_{b' \mathbf k' m'}^*(\mathbf r)
            \, \nabla
            \psi_{b \mathbf k m}(\mathbf r)
            -
            \right. \nonumber \nonumber \\
&&          \hspace{-5mm}
            \left.
            \psi_{b \mathbf k m}(\mathbf r)
            \, \nabla
            \psi_{b' \mathbf k' m'}^*(\mathbf r)
        \right]
        \, {a}_{b' \mathbf k' m'}^\dag(t)\,
        {a}_{b \mathbf k m}(t)\,,
\label{eq:jparamag}
\end{eqnarray}
and for the diamagnetic term:
\begin{eqnarray} \label{Eq:DiaCurrent}
    \hat{\mathbf j}^{(d)}(\mathbf r,t)
&=& -\,\frac{q}{m_e}
    \mathbf A(\mathbf r, t)
    \sum_{\scriptsize
                \begin{array}{c}
                b' \mathbf k' m'  \\
                b  \mathbf k  m
              \end{array}
             }
    \psi_{b' \mathbf k' m'}^*(\mathbf r) \psi_{b \mathbf k m}(\mathbf r)
    \times \nonumber\nonumber \\
&& \hspace{-5mm}
     {a}_{b' \mathbf k' m'}^\dag(t)\,
    {a}_{b \mathbf k m}(t)\,.
\label{eq:jdiamag}
\end{eqnarray}
In what follows both contributions will be studied in detail.
The simplification of the expressions will proceed in a similar manner
to the calculation of the transferred angular momentum.
However, at a certain point we will make use of a space average
($\overline{A} = (1/a^3) \int_{cell}\,d^3r\,A$)
in order to eliminate irrelevant microscopic (intra-cell) details.

% ---------------------------------------------------------------------------------
\subsubsection{Paramagnetic-current density}

Now we work out the expression (\ref{eq:jparamag}) by replacing the
wave functions from Eq.\ (\ref{eq:electron_basis}).
We apply the gradient operator on the envelope and periodic parts of
the Bloch wave function,
perform space average,
take mean value over the initial state,
and
split the result into interband-coherence and
population and intraband-coherence contributions.

\begin{figure}[h]
  \center\includegraphics[scale=0.8]{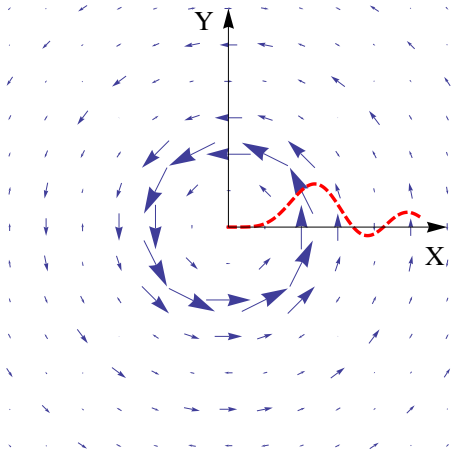}
  \caption{(color online) First-order paramagnetic current for light
            with $l=1$ and polarization $\sigma-$.
            The center of the plot coincides with the beam axis.
            The factor $J_{m-l}(k_r' r) J_{m}(k_r r)$ is drawn in dashed
            (red) line.
  \label{Fig:J}}
\end{figure}

%------------------------------------------------------------------------------
{\it Interband-coherence contribution:}
The interband-coherence contribution to the current density is given by
\begin{eqnarray*}
    \overline{\mathbf j}^{(coh)}(\mathbf r, t)
&=&     \frac{2\,q}{m_e}
        \sum_{\scriptsize
                \begin{array}{c}
                \mathbf k' m' \\
                \mathbf k  m
              \end{array}
             }
        \Re\Big[
            \mathbf p_{vc} \, {\cal N}'{\cal N}
            J_{m'}(k_r' r) J_{m}(k_r r) \times
%            \right.
\\
&& \hspace{-15mm}
        \left.
            e^{i(k_z-k_z')z}
            e^{i(m-m')\phi}
        \rho_{v \mathbf k' m', c \mathbf k m}^{(1)}(t)
        \right]
        \,
        \,.
\end{eqnarray*}
With the help of Eqs.\ (\ref{eq:Matrix_Element}) and (\ref{Eq_lowestO_pol})
we simplify this expression to
\begin{eqnarray*}
    \overline{\mathbf j}^{(coh)}(\mathbf r, t)
&=& \frac{2\, q}{m_e}
        \Re\Bigg[
            \mathbf p_{vc} e^{i q_z z} e^{i l \phi}
        \sum_{\scriptsize
                \begin{array}{c}
                k_r' k_r \\
                k_z  m
              \end{array}
             }
             {\cal N}'{\cal N}
             \times
%             \right.
\\
&& \hspace{-15mm}
        \left.
        J_{m-l}(k_r' r) J_{m}(k_r r)
        \, \rho_{v k_z-q_z k_r' m-l, c k_z k_r m}^{(1)}(t)
        \right]
        \,.
\end{eqnarray*}
\begin{figure}[h]
  \center\includegraphics[scale=0.8]{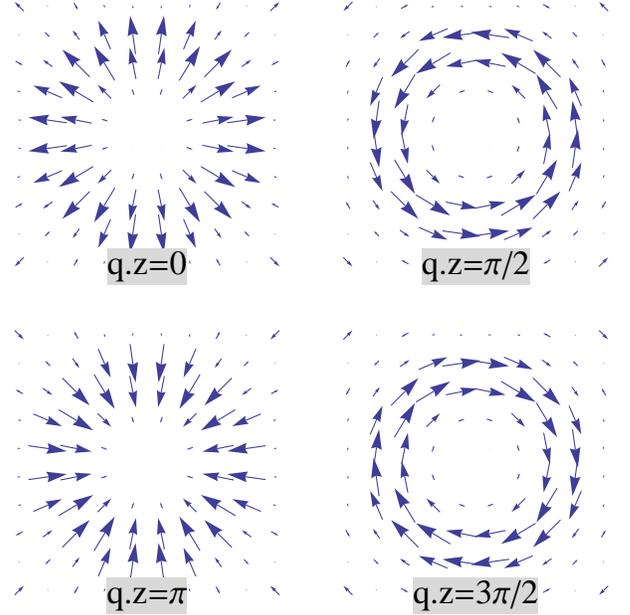}
  \caption{(color online) First-order paramagnetic current for light with
  $l=1$. The  center of each plot coincides with the beam axis.
  Each  panel pictures a different slice along the $z$-axis,
  showing the wave nature and space variation of the pattern.
  \label{Fig:Jwave}}
\end{figure}

\noindent
The main feature of this expression is that it contains a sum over
products of space- and time-dependent functions.
The time dependence presents two distinct scales, as mentioned before:
a rapid oscillation related to the frequency $\omega$ of the light beam,
and a slower one that is related to the detuning.

\begin{figure}[h]
  \center\includegraphics[scale=0.8]{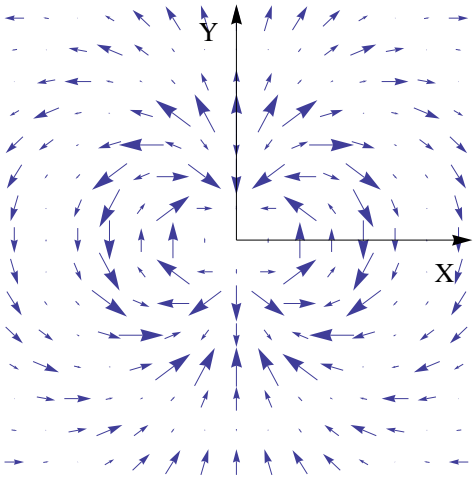}
  \caption{(color online) First-order paramagnetic current for light with
  $l=2$. The center of the plot coincides with the beam axis.
  The space pattern is more complex than in the $l=1$ case and exhibits
  circulation around off-axis centers.
  \label{Fig:Jlargel}}
\end{figure}

To describe the electric-current patterns in the plane perpendicular
to the $z$-axis, we disregard the slow time evolution and focus on
the space- and time-dependent quantity
$\Re[\mathbf p_{vc} e^{i (q_z z + l \phi - \omega t)}
J_{m-l}(k_r' r) J_{m}(k_r r)]$, see Fig.\ \ref{Fig:J}.
We can see that the current density and the angular momentum at this
level are consistent with each other.
To see this, let us take as an example a TL field having circular
polarization $\sigma_+ = \hat{x} + i \hat{y}$.
From the usual selection rule for the absorption of a photon, a
non-vanishing light-matter matrix element requires
$\mathbf p_{c v} = p_0(\hat{x} - i \hat{y})$.
Since $\mathbf p_{v c} = \mathbf p_{c v}^* = p_0(\hat{x} + i \hat{y})$,
we see from Eq.\ (\ref{eq:Lzcoh}) that a non-vanishing angular momentum
will appear only if $l=-1$;
in this case the current pattern reflects this fact, presenting
a ``circulation'' around the beam axis.
On the contrary, if the beam is tuned to $l=+1$ and $\sigma+$, the
current forms a pattern that does not ``flow'' around the axis.
As we move along the $z$-axis for fixed time, we observe the wave
nature of the x-y plane current (Fig.\ \ref{Fig:Jwave}).
For fixed $z$, the wave nature is revealed as time evolves.
For different values of $l$, the electric current develops more
complex patterns, which mimic the complex structure of the electric
field of a TL beam.
In the case of $|l|>1$, there appear more than one off-centered vortices,
as illustrated in Fig.\ \ref{Fig:Jlargel}.

From the above we can extract some general features of
$\overline{\mathbf j}^{(coh)}(\mathbf r, t)$.
Driven by the interband coherence, it oscillates in time, with zero mean,
at the frequency of the EM field, and presents complex spatial patterns that
are related to the peculiar electric field of the TL beam.
For special cases, the spatial pattern displays a circulation (of
microscopic origin and not to be confused with a macroscopic
excursion of the electrons around the beam axis) related to the
non-vanishing coherence contribution of angular momentum calculated in
Sec.\ \ref{Sec:OAM}.

%------------------------------------------------------------------------------
{\it Population and intraband-coherence contribution:}
The population and intraband-coherence contribution to the
paramagnetic current is given by
\begin{eqnarray*}
    \overline{\mathbf j}^{(pop)}(\mathbf r, t)
&=& - i\,\frac{q\,\hbar}{2\,m_e} \sum_{\scriptsize
                \begin{array}{c}
                 \mathbf k' m'  \\
                 \mathbf k m
              \end{array}
            }
        \left\{
            [\Phi_{\mathbf k' m'}^*(\mathbf r) \nabla
             \Phi_{\mathbf k m}(\mathbf r)] -
        \right.
\\
&&      \hspace{-18mm}
        \left.
           [\Phi_{\mathbf k m}(\mathbf r) \nabla
            \Phi_{\mathbf k' m'}^*(\mathbf r)]
        \right\}
        \, \rho_{c\mathbf k' m', \mathbf k m}^{(2)}(t)
        +
        \{ c \rightarrow v \}
        \,,
\end{eqnarray*}
where $\rho_{c\mathbf k' m', \mathbf k m}^{(2)}(t)$ is given by
Eqs.\ (\ref{Eq_lowestO_pop}) and (\ref{eq:prod_mat_elem}), and $\{ c
\rightarrow v \}$ stands for a similar term replacing $c$ by $v$.
The intraband current in the direction of $\hat{\phi}$ is
\begin{eqnarray}
    \overline{j}^{(pop)}_\phi(\mathbf r, t)
&=&   \frac{q\,\hbar}{m_e}
            \!\! \sum_{\scriptsize
                \begin{array}{c}
                  k_r' k_r  \\
                  m k_z
              \end{array}
         }
        {\cal N}'{\cal N}
         m \, \frac{1}{r}
            J_m(k_r' r) J_m(k_r r)
       \times \nonumber
\\
&&      \hspace{0mm}
       \rho_{c k_r' k_z m, k_r k_z m}^{(2)}(t)
        +
        \{c \rightarrow v \}
        \,.
\label{Eq:Jpop_transv}
\end{eqnarray}
Given that the electrons excited by the TL beam occupy initially a
portion of the valence band that is symmetric with respect to the
$\Gamma$ point of the Brillouin zone, we have disregarded the
contribution to $\overline{j}^{(pop)}_\phi(\mathbf r, t)$ coming
from the holes left behind, and kept only the current produced by
the imbalance of electrons in the conduction band.
The parameter $l$ does not appear explicitly, but it enters
$\rho_{c k_r' k_z m, k_r k_z m}^{(2)}(t)$, since an electron leaving a
state with $\{v,m\}$ goes to a state with $\{c,m+l\}$.

%-----------------------------------------------------------------------------
\subsubsection{Diamagnetic current density}

Starting from Eq.\ (\ref{Eq:DiaCurrent}), we perform a space average
and obtain for the diamagnetic term
\begin{eqnarray*}
    \mathbf j^{(d)}(\mathbf r,t)
&=& -\,\frac{q}{m_e}
    \mathbf A(\mathbf r, t)
    \sum_{\scriptsize
                \begin{array}{c}
                k_r' k_r  \\
                k_z m
              \end{array}
             }
    {\cal N}' {\cal N}
    J_m(k_r' r) J_m(k_r r)
        \times \nonumber \\
&& \hspace{-5mm}
     \rho_{c k_r' k_z m,  k_r k_z m}^{(2)}(t)
     +        \{c \rightarrow v \} \,,
\end{eqnarray*}
We point out the following features of the diamagnetic current
density: $i)$ it arises from populations and intraband coherences in
each band; $ii)$ its vectorial character is given by the
polarization of the light; $iii)$ it is of third order in the vector
potential amplitude; $iv)$ it does not arise from an imbalance of
the conduction-band population [it is not proportional to $m$ like
$\overline{j}^{(pop)}_\phi(\mathbf r, t)$, see Eq.\
(\ref{Eq:Jpop_transv})]; $v)$ its time evolution is given by the
slow evolution of $\rho_{c}^{(2)}(t)$ and the fast oscillation of
the field $\mathbf A(\mathbf r, t)$.

% ----------------------------------------------------------------
% ----------------------------------------------------------------
\section{Conclusions}
\label{sec:conc}

We developed a theory of
the band-to-band transitions induced by twisted light (light
carrying orbital angular momentum) in bulk semiconductors.
We posed the problem of the light-matter interaction in terms of
Heisenberg equations of motion for the populations and quantum coherences
of a two-band semiconductor model, as customarily done.
We found that the resulting system of equations is greatly simplified
when the envelope electron wave functions are represented in
cylindrical coordinates, instead of using the usual (cartesian) Bloch
state representation.
This simplification is due to the decoupling of the system of equations
in subsystems determined by the orbital angular momentum of electrons.
Though non-standard for the bulk case, our choice of basis states is,
on the one hand, perfectly admissible and, on the other, it proves to
be the best choice from a mathematical point of view.
It is admissible because the material properties are unaffected by surface
effects in the limit of a bulk/large system.
It is the right choice since it provides the highest symmetry
compatibility between the twisted-light vector potential and the
electron states.

Despite the achieved simplification, the evolution of the different relevant
physical quantities under the excitation by a time-dependent pulse must
be computed via numerical analysis of the equations of motion.
This task is left for future work; instead, we showed here analytical
results in the low-excitation or perturbative regime.
With the solutions for the populations and quantum coherences, we
confirmed on more solid grounds our previous findings, i.e.\ that the
optical excitation will generate electric currents, and that there
will be a transfer of orbital angular momentum from the light beam
to the electrons.
Our analysis of the electric current and electron's orbital angular momentum
showed that two qualitatively different contributions enter both
observables; they may be termed microscopic and macroscopic
contributions.
The microscopic contribution relates to the interband coherence and mimics
the behavior of the electric field; from this and other reasons, it
parallels the induced polarization of a semiconductor in the presence of
plane-waves, as traditionally studied using the vertical-transition
 assumption.
On the other hand, the macroscopic contribution signals a net transfer
of OAM from the field to the electrons, and it parallels the photon-drag
effect.
We showed that the electric current exhibits a high degree of spatial
complexity, due to the inhomogeneous nature of the twisted-light beam.
Additionally, we have calculated and briefly analyzed the diamagnetic
current, not addressed in our previous study.

\begin{acknowledgments}

We are grateful to Pierre Gilliot, Jamal Berakdar and Bj\"{o}rn
Sbierski for useful discussions.
We acknowledge support through grants ANPCyT PICT-2006-02134 and
UBACyT X495.

\end{acknowledgments}

% ----------------------------------------------------------------------------
% ----------------------------------------------------------------------------
% ----------------------------------------------------------------------------

\appendix

\section{Particle in a hollow cylinder}
\label{sec:cylinder_states}

We present the complete derivation of the electronic states in
cylindrical coordinates, starting from the well-known solution of
the Schr\"{o}dinger equation without potential
\begin{equation*}
-\frac{\hbar^2}{2m}\nabla^2\Phi(\mathbf r) = E\Phi(\mathbf r)\,,
\end{equation*}
where the Laplacian in cylindrical coordinates is
\begin{equation*}
    \nabla^2 f = \frac{1}{r} \frac{\partial}{\partial r}
  \left( r \frac{\partial f}{\partial r} \right)
+ \frac{1}{r^2} \frac{\partial^2 f}{\partial \theta^2} +
  \frac{\partial^2 f}{\partial z^2 }\,.
\end{equation*}
Consider a solution of the separable form $\Phi(\mathbf r) = R(r)
\Theta(\theta) Z(z)$; replacing into the Schr\"{o}dinger equation,
we get
\begin{equation*}
  -\frac{\hbar^2}{2m}
  \left[
    {1 \over R} {1 \over r} {\partial \over \partial r}
    \left( r {\partial R \over \partial r} \right) +
    {1 \over \Theta} {1 \over r^2}
    {\partial^2 \Theta \over \partial \theta^2} +
    {1 \over Z} {\partial^2 Z \over \partial z^2 }
  \right]
= E
\end{equation*}
which yields
\begin{eqnarray*}
    \left[
        {1 \over R} {1 \over r} {\partial \over \partial r}
        \left( r {\partial R \over \partial r} \right) +
        {1 \over Z} {\partial^2 Z \over \partial z^2 }
        \right] +
        {1 \over \Theta} {1 \over r^2}
        {\partial^2 \Theta \over \partial \theta^2}
&=& - \alpha^2
\end{eqnarray*}
where $\frac{2m}{\hbar^2} E = \alpha^2$. Then
\begin{eqnarray*}
    \left[
        {1 \over R} {r} {\partial \over \partial r}
        \left( r {\partial R \over \partial r} \right) +
        {r^2 \over Z} {\partial^2 Z \over \partial z^2 } +
        \alpha^2 r^2
    \right] +
        {1 \over \Theta}
        {\partial^2 \Theta \over \partial \theta^2}
&=& 0\,,
\end{eqnarray*}
which splits to
\begin{eqnarray*}
        {1 \over R} {r} {\partial \over \partial r}
        \left( r {\partial R \over \partial r} \right) +
        {r^2 \over Z} {\partial^2 Z \over \partial z^2 } +
        \alpha^2 r^2
&=& m^2 \\
        {1 \over \Theta}
        {\partial^2 \Theta \over \partial \theta^2}
&=& -m^2 \,.
\end{eqnarray*}
The second equation has solution $\Theta = A_1 e^{i m \theta} + A_2
e^{-i m \theta}$. The remaining equation is
\begin{eqnarray*}
    {1 \over R} {1 \over r} {\partial \over \partial r}
    \left( r {\partial R \over \partial r} \right) +
    \alpha^2 -
    {m^2 \over r^2} +
    {1 \over Z} {\partial^2 Z \over \partial z^2 }
&=& 0
        \,.
\end{eqnarray*}
This is again separable
\begin{eqnarray*}
    {1 \over R} {1 \over r} {\partial \over \partial r}
    \left( r {\partial R \over \partial r} \right) +
    \alpha^2 -
    {m^2 \over r^2}
&=& \lambda^2 \\
    {1 \over Z} {\partial^2 Z \over \partial z^2 }
&=& - \lambda^2
    \,,
\end{eqnarray*}
with solution $Z = B e^{i \lambda z}$, and equation
\begin{eqnarray*}
    {1 \over R}  r {\partial \over \partial r}
    \left( r {\partial R \over \partial r} \right) +
    (\alpha^2 - \lambda^2) r^2 -
    {m^2}
&=& 0
    \,,
\end{eqnarray*}
and developing the derivatives
\begin{eqnarray*}
    r^2 {d^2 R \over d r^2}  +
    r   {d R \over d r} +
    \left[
        (\alpha^2 - \lambda^2) r^2 - {m^2}
    \right] R
&=& 0
    \,,
\end{eqnarray*}
which is almost the Bessel differential equation.
Defining $x^2 = (\alpha^2 - \lambda^2) r^2$ the correct differential
equation follows
\begin{eqnarray*}
    x^2 {d^2 R \over d x^2}  +
    x   {d R \over d x} +
    \left(
        x^2 - {m^2}
    \right) R
&=& 0
    \,.
\end{eqnarray*}

% ----------------
\subsection{Boundary conditions}

For $Z(z)$ we use boundary conditions $Z(0)=Z(H)$, and obtain
\begin{eqnarray*}
    Z(z)
&=& {1 \over \sqrt{H}} \, e^{i \lambda z} \\
    \lambda
&=& {2 \pi n \over H}
\,.
\end{eqnarray*}
For $\Theta(\theta)$, the periodicity implies
\begin{eqnarray*}
    \Theta(\theta)
&=& N_\theta \, e^{i m \theta} \\
    m
&=& \ldots,-2,-1,0,1,2, \ldots
\,.
\end{eqnarray*}
For the radial solution, since it has to be finite at the origin,
the solutions are the Bessel functions of the first kind $J_m (x)$.
If we demand that $J_m (x=\sqrt{\alpha^2 - \lambda^2} L) = 0$, then
the $s$th zero $r_{ms}$ of the Bessel function should be
$ \alpha^2 = {(r_{m s}/L)^2 + \lambda^2}$:
\begin{eqnarray*}
    E_{nms}
&=& \frac{\hbar^2}{2 m} \left[ \left(\frac{r_{m s}}{L}\right)^2 +
    \left(\frac{2 \pi n}{H}\right)^2 \right] \\
    \Phi_{nms}(\mathbf r)
&=& {\cal N}_{ms} J_m(\sqrt{\alpha^2 - \lambda^2}\,r)
e^{i m \theta} e^{i {2 \pi n \over H} z} \,,
\end{eqnarray*}
with normalization
${\cal N}_{ms}=\frac{2}{L J_{m}'(r_{ms})}\sqrt{\frac{1}{2 \pi \, H }}$,
and $\lambda < \alpha$ for any admissible solution.

% --------------------------------------------------------------
% --------------------------------------------------------------
\subsection{Cylinder with a Bravais lattice}

In the effective-mass approximation, the complete wave function is
expressed as the product of an envelope $\Phi_{nms}(\mathbf r)$ and
a periodic $u(\mathbf r)$ function.
Then, the Schr\"{o}dinger equation reads
%
%\begin{widetext}
\begin{eqnarray*}
    -\frac{\hbar^2}{2m}u(\mathbf r) \nabla^2
    \Phi_{nms}(\mathbf r) -
    \frac{\hbar^2}{2m} \Phi_{nms}(\mathbf r)
    \nabla^2  u(\mathbf r) - \\
    \frac{\hbar^2}{m} [\nabla \Phi_{nms}(\mathbf r)] \cdot
    [\nabla u(\mathbf r)] +
    U(\mathbf r) [\Phi_{nms}(\mathbf r)u(\mathbf r)] = \\
&& \hspace{-20mm}
    E\Phi(\mathbf r) u(\mathbf r)
\,,
\end{eqnarray*}
%\end{widetext}
%
where $U(\mathbf r)$ is the lattice potential. Since we have already
solved the problem of the free particle without $u(\mathbf r)$, we
know that $-\frac{\hbar^2}{2m} \nabla^2 \Phi_{nms}(\mathbf r) =
E^{(0)}_{nms} \Phi_{nms}(\mathbf r)$.
Dividing by $\Phi_{nms}(\mathbf r)$ and grouping terms we get
\begin{eqnarray*}
    - \frac{\hbar^2}{2m} \nabla^2  u(\mathbf r) +
    U(\mathbf r) u(\mathbf r) - \frac{\hbar^2}{m}
    \frac{\nabla \Phi_{nms}(\mathbf r) \cdot
    \nabla u(\mathbf r)}{\Phi_{nms}(\mathbf r) } && \\
=   (E - E^{(0)}_{nms}) u(\mathbf r) \,. &&
\end{eqnarray*}
This is an equation for $ u(\mathbf r)$ [since we already know the
functional form of $\Phi_{nms}(\mathbf r)$].
We ask that $u(\mathbf r) =  u(\mathbf r + \mathbf R)$,
with $\mathbf R$ a lattice vector, so we can regard the region of
integration as the unit cell.
Then, if $\Phi_{nms}$ varies slowly in the unit cell, we see
$(\hbar^2/m)
    \Phi_{nms}(\mathbf r) ^{-1}
    [\nabla \Phi_{nms}(\mathbf r)] \cdot
    [\nabla u(\mathbf r)] $ as a perturbation.
This is simply the $\bf k \cdot p$ approximation in a different
coordinate system.
To lowest order we have
\begin{eqnarray*}
    - \frac{\hbar^2}{2m} \nabla^2  u(\mathbf r) +
    U(\mathbf r) u(\mathbf r)
&=& \Delta E \, u(\mathbf r)
\,,
\end{eqnarray*}
and from here one obtains $u(\mathbf r)$ as usual.
Therefore, we take the eigenfunctions as a product of envelope and
periodic functions, while the energy is that of the envelope
corrected by the energy gap and the effective mass:
\begin{eqnarray*}
    E_{bnms}
&=& \frac{\hbar^2}{2 m_b^*}
    \left[ \left(\frac{r_{m s}}{L}\right)^2 +
    \left(\frac{2 \pi n}{H}\right)^2 \right] + \delta_{b c} E_g
    \nonumber \\
    \Psi_{bnms}(\mathbf r)
&=& \frac{ N_r}{\sqrt{2 \pi H}} J_m[({r_{ms} / L})\,r]
    e^{i m \theta} e^{i {2 \pi n \over H} z} u_{b}(\mathbf r)
\,,
\label{eq:electron_basis_app}
\end{eqnarray*}
where $H$ is the height and $L$ the radius of the cylinder.

% ---------------------------

\end{document}